\documentclass[%
 reprint,
 amsmath,amssymb,
 aps,
superscriptaddress,
prl,
floatfix,
]{revtex4-2}

\usepackage{balance}
\usepackage{xcolor}
\usepackage{graphicx}
\usepackage{dcolumn}
\usepackage{bm}

\newcommand{\Yong}[1]{{\color{black}#1}}

\newcommand{\Andrew}[1]{{\color{black}#1}}

\newcommand{\Fedosejevs}[1]{{\color{black}#1}}

\newcommand{\Peter}[1]{{\color{black}#1}}

\newcommand{\John}[1]{{\color{black}#1}}

\begin{document}

\preprint{APS/123-QED}

\title{Electron Dynamics in a Wakefield Accelerator Driven by \\Laser Pulses Carrying \Yong{Orbital Angular Momentum}}

\author{Q. Qian}
\email[Corresponding author: ]{qqbruce@umich.edu}
\affiliation{G\'{e}rard Mourou Center for Ultrafast Optical Science, University of Michigan, 2200 Bonisteel Boulevard, Ann Arbor, Michigan 48109, USA}
\author{M. Li}
\email[Corresponding author: ]{molix@umich.edu }
\thanks{This author is now at State Key Laboratory of Dark Matter Physics, Key Laboratory for Laser Plasmas (MOE), School of Physics and Astronomy, Shanghai Jiao Tong University, 200240 Shanghai, China}
\affiliation{G\'{e}rard Mourou Center for Ultrafast Optical Science, University of Michigan, 2200 Bonisteel Boulevard, Ann Arbor, Michigan 48109, USA}
\author{Y. Ma}
\affiliation{G\'{e}rard Mourou Center for Ultrafast Optical Science, University of Michigan, 2200 Bonisteel Boulevard, Ann Arbor, Michigan 48109, USA}
\author{P. T. Campbell}
\affiliation{G\'{e}rard Mourou Center for Ultrafast Optical Science, University of Michigan, 2200 Bonisteel Boulevard, Ann Arbor, Michigan 48109, USA}
\author{C-H. Chang}
\affiliation{G\'{e}rard Mourou Center for Ultrafast Optical Science, University of Michigan, 2200 Bonisteel Boulevard, Ann Arbor, Michigan 48109, USA}
\author{E. Denis}
\affiliation{Atomic and Laser Physics Sub-Department Clarendon Laboratory, Department of Physics, University of Oxford, Oxford, UK}
\author{N. Ernst}
\affiliation{G\'{e}rard Mourou Center for Ultrafast Optical Science, University of Michigan, 2200 Bonisteel Boulevard, Ann Arbor, Michigan 48109, USA}
\author{R. Fedosejevs}
\affiliation{Department of Electrical and Computer Engineering, University of Alberta, Edmonton, T6G1R1, Canada}
\author{B. Hou}
\affiliation{G\'{e}rard Mourou Center for Ultrafast Optical Science, University of Michigan, 2200 Bonisteel Boulevard, Ann Arbor, Michigan 48109, USA}
\author{A. James}
\affiliation{Atomic and Laser Physics Sub-Department Clarendon Laboratory, Department of Physics, University of Oxford, Oxford, UK}
\author{M. W. von der Leyen}
\affiliation{Atomic and Laser Physics Sub-Department Clarendon Laboratory, Department of Physics, University of Oxford, Oxford, UK}
\author{K. Krushelnick}
\affiliation{G\'{e}rard Mourou Center for Ultrafast Optical Science, University of Michigan, 2200 Bonisteel Boulevard, Ann Arbor, Michigan 48109, USA}
\author{C. Kuranz}
\affiliation{G\'{e}rard Mourou Center for Ultrafast Optical Science, University of Michigan, 2200 Bonisteel Boulevard, Ann Arbor, Michigan 48109, USA}
\author{A. Longman}
\affiliation{Lawrence Livermore National Laboratory, 7000 East Avenue, Livermore, California 94550, USA}
\author{A. Maksimchuk}
\affiliation{G\'{e}rard Mourou Center for Ultrafast Optical Science, University of Michigan, 2200 Bonisteel Boulevard, Ann Arbor, Michigan 48109, USA}
\author{J. Nees}
\affiliation{G\'{e}rard Mourou Center for Ultrafast Optical Science, University of Michigan, 2200 Bonisteel Boulevard, Ann Arbor, Michigan 48109, USA}
\author{P. Norreys}
\affiliation{Atomic and Laser Physics Sub-Department Clarendon Laboratory, Department of Physics, University of Oxford, Oxford, UK}
\affiliation{John Adams Institute for Accelerator Science, Department of Physics, University of Oxford, Oxford, UK}
\author{T. Nutting}
\affiliation{G\'{e}rard Mourou Center for Ultrafast Optical Science, University of Michigan, 2200 Bonisteel Boulevard, Ann Arbor, Michigan 48109, USA}
\author{J. B. Ohland}
\affiliation{GSI Helmholtzzentrum für Schwerionenforschung, Planckstraße 1, 64291 Darmstadt, Germany}
\author{J. P. Palastro}
\affiliation{Laboratory for Laser Energetics, University of Rochester, Rochester, New York 14623, USA}
\author{A. G. R. Thomas}
\affiliation{G\'{e}rard Mourou Center for Ultrafast Optical Science, University of Michigan, 2200 Bonisteel Boulevard, Ann Arbor, Michigan 48109, USA}
\author{R. Timmis}
\affiliation{Atomic and Laser Physics Sub-Department, Clarendon Laboratory, Department of Physics, University of Oxford, Oxford, UK}
\affiliation{John Adams Institute for Accelerator Science, Department of Physics, University of Oxford, Oxford, UK}

\author{L. Willingale}
\affiliation{G\'{e}rard Mourou Center for Ultrafast Optical Science, University of Michigan, 2200 Bonisteel Boulevard, Ann Arbor, Michigan 48109, USA}
\author{M. Burger}
\email[Corresponding author: ]{milosb@umich.edu}
\affiliation{G\'{e}rard Mourou Center for Ultrafast Optical Science, University of Michigan, 2200 Bonisteel Boulevard, Ann Arbor, Michigan 48109, USA}




\date{\today}

\begin{abstract}
Laser pulses carrying orbital angular momentum (OAM) provide a new degree of freedom for controlling plasma-based accelerators. Here, we experimentally demonstrate OAM-driven laser wakefield acceleration, producing electron beams with a two-beamlet structure that indicates unique azimuthal dynamics inside the plasma wake. Particle-in-cell simulations reproduced the observed spectral features and revealed helical electron trajectories driven by OAM-pulse-driven wakefields. These results show that the laser driver's phase structure can shape electron acceleration dynamics, opening a route to optimal control of beam structure in compact laser-driven accelerators.  
\end{abstract}

\maketitle




Laser-wakefield acceleration (LWFA) \Yong{~\cite{Tajima79, Mangles04, Geddes04}} enables the production of \Yong{high-quality} electron beams over millimeter-scale distances by driving plasma waves with ultrashort laser pulses. In the nonlinear blowout regime~\cite{Lu06}, the plasma wake structure depends on the spatiotemporal characteristics of the drive laser. Substantial efforts have been devoted to controlling beam quality by manipulating laser intensity, pulse duration, plasma density, and ionization dynamics~\cite{Esarey09}. Despite these advances, control over the transverse phase structure of the driver laser pulse has remained relatively unexplored as a mechanism for shaping acceleration dynamics.

Laser pulses carrying orbital angular momentum (OAM), most commonly realized as Laguerre–Gaussian (LG) modes~\cite{Allen92}, introduce \Andrew{qualitatively} a new degree of freedom in laser–plasma interactions. The LG field can be written as
\begin{equation}
    \tilde{E}_{mp}
    = a_{0,\mathrm{LG}} f_{mp}Ge^{-im\Phi},
\end{equation}
where $a_{0,\mathrm{LG}}$ is the energy-normalized amplitude parameter, $G$ is the Gaussian distribution; $f_{mp}$ is the Laguerre polynomial; and $\exp(im\Phi)$ is the helical phase front, with $m$ representing the topological charge or the LG azimuthal index. The OAM pulse features \John{a central singularity in the phase with a distinctive annular intensity profile}. Consequently, the ponderomotive force of such OAM pulses differs fundamentally from that of a conventional Gaussian \John{pulse}. An OAM pulse drives an annular, rather than spherical, blowout cavity. \Peter{Notably, theoretical and numerical studies have shown that structured wakefields are particularly attractive for positron acceleration, offering accelerating gradients comparable to those for electrons while providing transverse focusing~\cite{Vieira14}. This capability is relevant for future high-energy-density applications such as the hybrid asymmetric linear Higgs Factory concept~\cite{Foster_NJP_2023}, in which a plasma-wakefield accelerator drives the electron arm while a conventional RF accelerator handles the positron beam.} Subsequent studies have further shown that such drivers can produce hollow or transversely structured electron beams and transfer angular momentum to accelerated particles~\cite{Zhang16, Wu_NJP_2019, Ju_NJP_2018, Shi_PPCF_2021}. When using multiple OAM pulses to create a helical spatiotemporal intensity profile, also known as a light spring\Yong{~\cite{Pariente_OL_15, Longman26}}, twisted plasma accelerators capable of generating relativistic \Yong{vortex electron} beams with helical motion can be realized. \John{In addition}, they can greatly enhance the energy and intensity of the radiated \John{X-rays}~\cite{Mendonca14, Vieira18, Mendonca22, Martins19}. Although these phenomena have been extensively investigated in particle-in-cell (PIC) simulations, direct experimental observations of OAM-induced modifications~\cite{Longman19} in the electron energy spectrum \John{have until now been lacking.}

Here, we report the experimental demonstration of laser wakefield acceleration driven by OAM laser pulses. \John{In the experiment, we observed the emergence of multi-beamlet electron energy spectra that exhibit a clear angular structure.} Compared to the single on-axis quasi-monoenergetic beam \John{characteristic of} Gaussian-driven LWFA, the OAM configuration yields two separated beamlets in the transverse momentum space. PIC simulations reproduced this behavior, indicating azimuthally distinct electron dynamics within a rotating wake that produce multiple electron bunches that remain spatially separated during propagation. 

\begin{figure}
	\centering  \includegraphics[width=\linewidth]{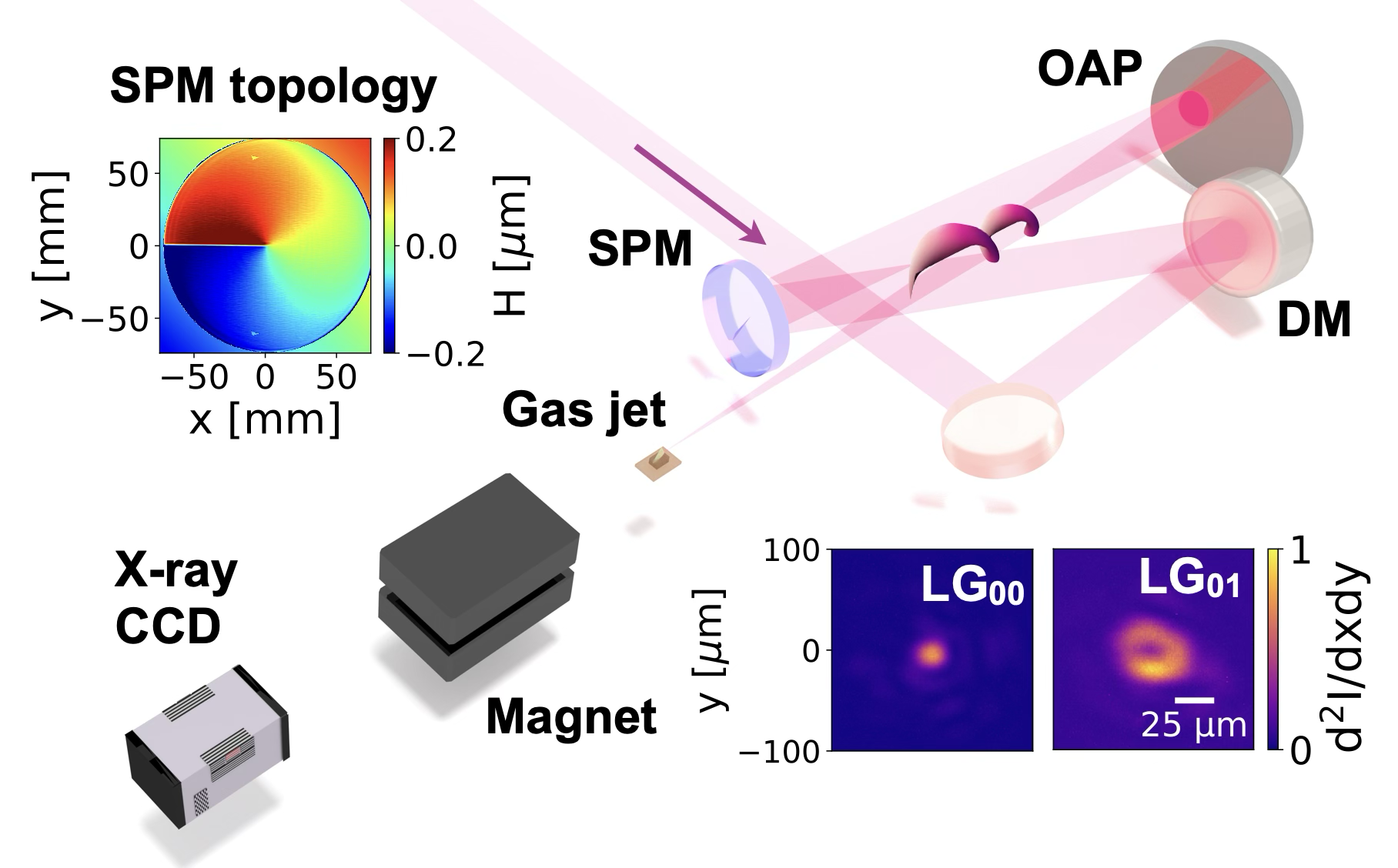}
\caption{Schematic of the experimental setup: DM - deformable mirror; SPM - spiral phase mirror; OAP - on-axis parabolic mirror. Insets show the surface height profile of the SPM and the focal spots of the Gaussian and OAM beams, normalized to their maximum intensity.}
\label{fig:Setup}
\end{figure}

\Yong{The experiment was carried out in Target Area 3 \John{of} the ZEUS Laser Facility at the University of Michigan~\cite{Maksimchuk25}}. Figure~\ref{fig:Setup} shows the setup. The beamline delivered an average \John{peak} energy of 3.73 $\text{J}$ to the target. The \Yong{full-width at half-maximum (FWHM) pulse duration was 29.4 $\text{fs}$}, resulting in a peak power of 126.87 $\text{TW}$. \Andrew{The 83.3-mm diameter beam was focused using a 2-m focal length spherical mirror, \John{with an approximate f-number of 24.}} \Andrew{OAM was introduced into the near-field using a 6$^\circ$ angle of incidence off-axis spiral phase mirror centered at 800 nm~\cite{Longman20} (see inset figure in Fig.\ref{fig:Setup}). \John{The mirror could be removed between shots, allowing for a direct comparison between} a Gaussian focal spot with no topological charge and a Laguerre-Gaussian focal spot with topological charge $m$ = 1. }
\Andrew{Example focal spot intensity distributions are shown in the inset of Fig.~\ref{fig:Setup}}. For the Gaussian LWFA, the FWHM of the laser focal spot is 23.5 $\mu$m by 25.1 $\mu$m. This results in an intensity of $1.9\times10^{19}$ $\text{W/cm}^2$ and a corresponding \John{peak} normalized vector potential $a_0\approx$ 2.9. \Andrew{Inserting the SPM converts the laser to an $m$ = 1 optical vortex, producing an annular intensity at the focus. The energy and pulse duration remain unchanged for both the Gaussian and OAM cases.
} 
\Andrew{The laser was linearly polarized in the horizontal direction and focused at the beginning of a 1-cm-long supersonic gas jet \cite{Yong_PRL_2024} composed of 97\% He and 3\% N$_2$, enabling ionization injection \Yong{\cite{Pak_PRL_2010, McGuffey_PRL_2010}}. The plasma density was controlled by the gas jet backing pressure and optimized for maximum electron energy, which was similar for the Gaussian and OAM cases. It was found to be approximately  $3\times10^{18} \text{cm}^{-3}$. \Yong{The density was measured with the Shearing Mach-Zehnder interferometer}.}

\begin{figure}
	\centering
\includegraphics[width=\linewidth]{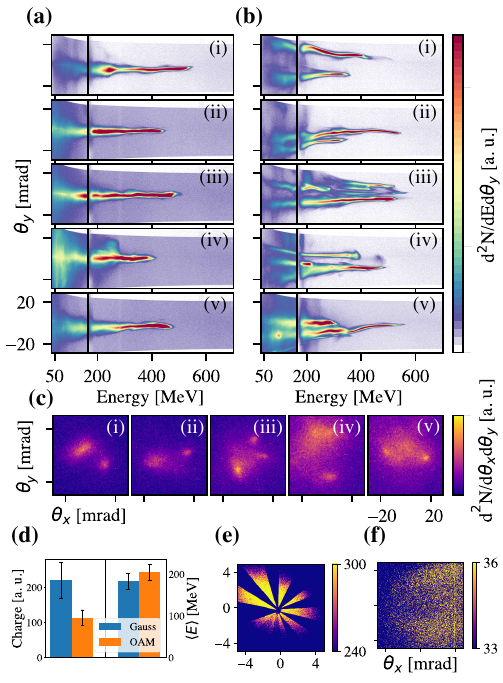}
\caption{(a), (b): Electron beam energy spectra generated by (a) Gaussian and (b) OAM (LG$_{01}$) laser \John{pulse}-driven wakefield. The electron beam from the Gaussian driver \John{diverges from the axis at a small angle}, whereas the OAM driver produces a two-beamlet structure. (c) Electron beam profiles for OAM pulse-driven wakefield. The beam profile shows a two-beamlet structure, and the separation of the two beamlets is about 20 to 40$~\mathrm{mrad}$. (d) The comparison of relative charge and mean energy for two cases. (e), (f) Betatron radiation on X-ray CCD for (e) Gaussian and (f) OAM pulse. An X-ray filter wheel was placed in front of the X-ray CCD for the Gaussian case \Andrew{but not for the OAM case because the signal was much weaker.}} 
\label{fig:spectrum_gauss_oam}
\end{figure}

Figure~\ref {fig:spectrum_gauss_oam} displays \Andrew{a subset of energy spectra measured for the Gaussian-driven LWFA (a) and OAM beam-driven LWFA (b).} In both cases, 300--550 MeV electron beams were generated. Compared with well-centered pencil-like energy spectra from Gaussian LWFA, the OAM LWFA beams have a unique two-beamlet structure. Each spectrum contains two \Yong{beamlets separated} along the spectrometer's vertical direction, which in our setup is perpendicular to the laser polarization direction. The separation between the two beamlets varies up to 40$~\mathrm {mrad} $, while the divergence and energy of the two beamlets are comparable.    


\Andrew{A scintillator screen was placed 18.8 cm downstream from the gas jet to measure the transverse spatial profile of the accelerated electron beams}. \Andrew{Example transverse spatial profiles are shown in Fig.~\ref{fig:spectrum_gauss_oam} (c) for the OAM pulse-driven cases.} \Andrew{We see that the OAM pulse case produces an electron beam with a bimodal distribution, and its orientation changes shot to shot.} \Andrew{The separation of the beams is on the order of 20-40 mrad, much larger than the $\approx$ 5 mrad FWHM of the individual beamlets themselves, indicating that the electron beams are being ejected from the interaction with an angle relative to the central laser $k$-vector.} 
The relative charge of the electrons in the OAM pulse-driven wakefield is about half that of the Gaussian beam-driven wakefield, and the averaged energy is about the same, as shown in Fig.~\ref{fig:spectrum_gauss_oam}(d). However, the betatron radiation intensity measured on the X-ray CCD is much lower compared to the Gaussian beam case, as shown in Fig.~\ref{fig:spectrum_gauss_oam} (e) and (f). For the Gaussian beam-driven case in Fig.~\ref{fig:spectrum_gauss_oam} (e), the filter wheel was placed in front of the X-ray CCD, and we observe a well-defined elliptical X-ray beam; for the OAM beam-driven case in Fig.~\ref {fig:spectrum_gauss_oam} (f), the signal level is much weaker, requiring the removal of the filter wheel to resolve the X-ray signal. This significant difference in X-ray intensity could be due to larger divergence. Still, it may also reflect differences in electron transverse dynamics in the wake between Gaussian- and OAM beam-driven LWFA, which warrants further investigation in future work.


To \Andrew{quantify} the features of the observed split energy spectrum and the characteristics of the electron beam profile, we performed the Fourier-Bessel particle-in-cell (FBPIC, \cite{Lehe16}) simulations with an OAM (LG$_{01}$) pulse having energy of
$U_L = 3.73$ J and an energy-normalized amplitude parameter \(a_{0,\mathrm{LG}}=2.9\). The waist parameter is \(w_0=20.4~\mu\mathrm{m}\), corresponding to a Gaussian intensity FWHM diameter \(2w_0\sqrt{\ln2/2}=24~\mu\mathrm{m}\), and the same temporal envelope parameter \(\tau=25~\mathrm{fs}\), corresponding to an intensity FWHM duration \(2\tau\sqrt{\ln2/2}=29.4~\mathrm{fs}\). The plasma electron density was \(3.0\times10^{18}~\mathrm{cm}^{-3}\) over a total interaction length of \(1~\mathrm{cm}\). The gas mixture consisted of \(97\%\) helium and \(3\%\) nitrogen; we studied three different configurations: ($i$) azimuthal mode $m$ = 1, with the laser polarized along the $\hat x$ direction; ($ii$) $m$ = 1, with the laser polarized along the $\hat y$ direction; and ($iii$) $m$ = -1, with the laser polarized along the $\hat x$ direction.


\begin{figure}
\centering
\includegraphics[width=\linewidth]{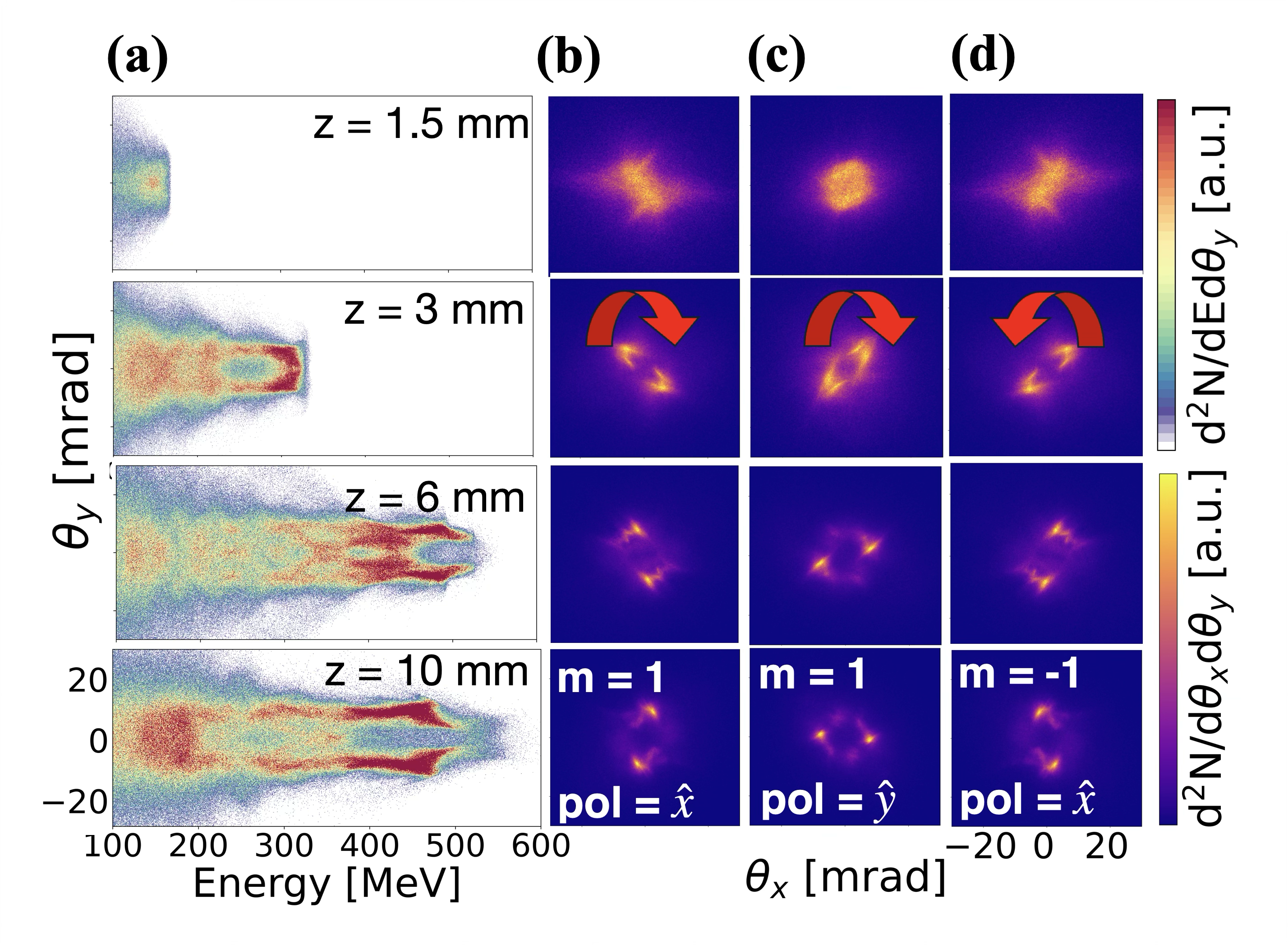}
\caption{The evolution of (a) the electron energy spectrum from the simulation. A two-beamlet structure appears at later times in the simulation. (b), (c), (d) The evolution of the electron beam profile with: (b) topological charge $m$ = 1, laser polarized along the $\hat x$ direction; (c) topological charge $m$ = 1, laser polarized along the $\hat y$ direction; and (d) topological charge $m$ = -1, laser polarized along the $\hat x$ direction.}
\label{fig:Espec_Eprofile_main}
\end{figure}

The simulated electron spectra and electron-profile distributions are shown in Fig.~\ref{fig:Espec_Eprofile_main}. The OAM pulse-driven LWFA exhibited a clear split structure perpendicular to the laser polarization direction in all three cases. Figure~\ref{fig:Espec_Eprofile_main} (a) shows the energy spectrum for $m$ = 1, and the laser is polarized along the $ \hat {x} $ direction. Since the electron spectrometer screen in the experiment was also placed perpendicular to the laser polarization direction, the simulated split provides a direct comparison with the experimentally observed split spectrum. The electron-profile diagnostic provides a consistent picture. The symmetric LG$_{01}$ case showed a bimodal distribution. The orientation of these bimodal distributions changes with the change in laser polarization direction, as shown in Fig.~\ref{fig:Espec_Eprofile_main} (b), (c). As the wakefield propagates, the two beamlets rotate, and the rotation direction correlates with the sign of the topological charge carried by the OAM pulse. For the $m$ = 1 case, the beam rotates clockwise, and for the $m$ = -1 case, the beam rotates counterclockwise, as shown in Fig.~\ref{fig:Espec_Eprofile_main} (b), (d).

\begin{figure}
\centering
\includegraphics[width=0.45\textwidth]{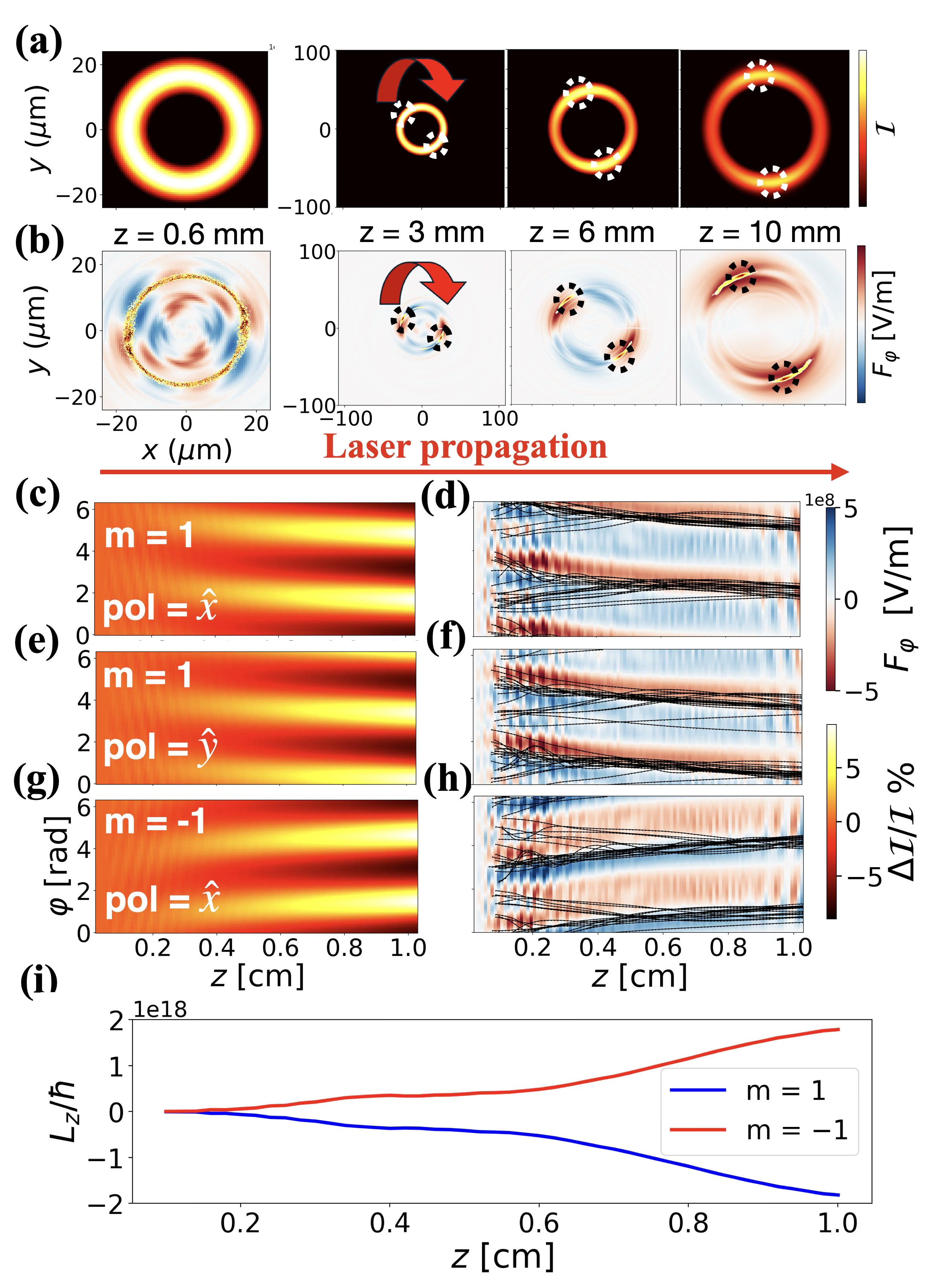}
\caption{The evolution of (a) the laser beam profile and (b) the azimuthal Lorentz force of the wakefield where the electron beam is located; the electron beam distribution is overlaid on top of the azimuthal force. (c), (e), (g): The evolution of the intensity perturbation along the azimuthal direction of the OAM beam inside the plasma. (d), (f), (h): The evolution of the azimuthal Lorentz force in the wake along the azimuthal direction. The overlapping black dotted curves represent the electron trajectories. (c), (d) $m$ = 1 with laser polarized along $\hat x$ direction; (e), (f) $m$ = 1 with laser polarized along $\hat y$ direction; (g), (h) $m$ = -1 with laser polarized along $\hat x$ direction. (i) The evolution of angular momentum carried by the electron beam.}
\label{fig:laser_wake_electron_dynamics}
\end{figure}

To further understand the transverse dynamics of the electron beam inside OAM-driven LWFA, we explore the evolution of the OAM beam intensity distribution, as well as the transverse wakefield along the azimuthal direction, averaged over the longitudinal span of the electron beams, as shown in Fig.~\ref{fig:laser_wake_electron_dynamics} (a) and (b). For an ideal donut-shaped wake driven by an OAM beam with a perfect ring shape, the transverse wakefield will only have a radial component due to symmetry. However, as this relativistically intense, linearly polarized OAM pulse propagates in plasma, intensity perturbations build up, and we observe two intense lobes gradually forming along the azimuthal direction, separated by 180$^\circ$. This intensity perturbation breaks the azimuthal symmetry and induces a wakefield along the azimuthal direction. This azimuthal field structure has two attraction points, which will trap \John{electrons, forming two beamlets} in the azimuthal direction, as shown in Fig. ~\ref{fig:laser_wake_electron_dynamics} (b), where the electron density distribution overlaps with the azimuthal wakefield distribution. From both Figs.~\ref{fig:laser_wake_electron_dynamics} (a) and (b), we can see that as the laser propagates, the azimuthal intensity perturbation for the OAM pulse rotates in the clockwise direction, and so does the azimuthal wakefield distribution. The rotation of the azimuthal wakefield then pushes the two beamlets to rotate and induces the bimodal distribution in the beam profile shown in Fig.~\ref{fig:Espec_Eprofile_main} (b) and (c).

Figures ~\ref{fig:laser_wake_electron_dynamics} (c)-(h) demonstrate the evolution of the OAM beam intensity perturbation and the transverse wakefield along the azimuthal direction, as well as the rotation of the intensity perturbation and the wakefield distribution as the beam propagates inside the plasma. Overlaid on top of Fig.~\ref{fig:laser_wake_electron_dynamics} (d), (f) and (h) are the electron trajectories, showing that the electrons are initially distributed uniformly along the azimuthal direction but get trapped into two phases as the laser propagates; the slope of the beam intensity modulation and azimuthal field distribution, as well as the electron trajectory, demonstrates the rotation in the azimuthal direction. The distribution gets an almost $\pi/2$ phase shift when changing the laser polarization from $\hat x$ to $\hat y$, as can be seen by comparing Fig.~\ref{fig:laser_wake_electron_dynamics} (c), (d) and (e), (f), which agrees with the beam profile shown in Figs.~\ref{fig:Espec_Eprofile_main} (b) and (c). The rotation direction, represented by the negative or positive slope in Fig.~\ref{fig:laser_wake_electron_dynamics} (c)-(h), can be controlled by changing the sign of the topological charge; in (c), (d) and (e), (f), $m$ = 1 case rotation is in the clockwise direction, while in (g) and (h) $m$ = -1 case, rotation is in the counterclockwise direction, which also agrees with the rotation of the beam profile shown in Figs.~\ref{fig:Espec_Eprofile_main} (b) and (d).

Finally, as a clear rotational motion of the beam can be \John{observed in the} simulation, we calculated the total mechanical angular momentum of the high-energy electron population (\(E \gtrsim 100~\mathrm{MeV}\)) about the laser propagation axis, defined as \(L_{z,\mathrm{tot}}=\sum_i w_i L_{z, i}\), where \(L_{z, i}=x_i p_{y, i}-y_i p_{x, i}\) and \(w_i\) is the macroparticle weight. Figure~\ref{fig:laser_wake_electron_dynamics} (i) shows the evolution of \(L_{z,\mathrm{tot}}\). The sign of the angular momentum carried by the electron changes with the sign of the topological charge of the OAM beam: for the $m$ = 1 case, \(L_{z,\mathrm{tot}}\) decreases monotonically from near zero to approximately \(-1.8\times10^{18}\hbar\); for the $m$ = -1 case, \(L_{z,\mathrm{tot}}\) increases monotonically from near zero to approximately \(1.8\times10^{18}\hbar\). This change in total angular momentum indicates that the OAM-driven wake will transfer OAM from the laser to the \John{trapped} electron beam.

The simulations indicate that the unique azimuthal dynamics of electrons inside OAM-driven wakefield produce beamlets with finite relative angular separation. Projecting these beamlets onto the spectrometer's vertical direction then produces the observed split traces. Shot-to-shot variations in the measured separation can be attributed to changes in the relative charge, orientation, and transverse momentum of the beamlets. {We distinguish the OAM-induced two-beamlet structure from conventional filamentation through several key features \cite{Yong_POP_2015, Elle_2018_NJP, Liu13}. While filamentation and related instabilities can produce multiple irregular beamlets, the present structure consistently features two dominant beamlets, as reproduced in simulations using an ideal symmetric LG$_{01}$ driver (Fig.~\ref{fig:Espec_Eprofile_main}). 
The systematic angular momentum transfer to the electrons (Fig.~\ref{fig:laser_wake_electron_dynamics} (i)) further supports the role of helical wake dynamics over stochastic filamentation.}

\Yong{In summary, we experimentally} \John{demonstrated} an LWFA driven by OAM laser pulses. The generated electron beams exhibit two distinct \Fedosejevs{spatial} branches that are well separated in the transverse momentum space. Complementary PIC simulations reproduced these features and showed that the spectral splitting arises from the electron azimuthal dynamics inside a near ring-shaped, rotating wakefield. The azimuthal trapping inside the wake forms spatially separated electron bunches that co-propagate and co-evolve during acceleration, and the OAM beam transfers angular momentum to the accelerated beam. These results demonstrate that the phase structure of the driving laser pulse can actively control electron transverse dynamics, enabling the generation of structured relativistic electron beams with tailored phase-space properties. The demonstrated electron azimuthal dynamics inside the wake may provide a pathway toward compact sources of structured X-ray radiation with orbital angular momentum, generated directly from laser-driven relativistic electron beams. 
\begin{acknowledgments}
This work was completed with support from the Department of Energy (DOE) under the award number DE-NA0002723. The construction and operation of the ZEUS facility is supported by the National Science Foundation (NSF) under Award No. 1935950 and 2126181, as well as by the AFOSR Grant No. FA9550-22-1-0118 and the University of Michigan. \Peter{The Oxford team is supported by UKRI-STFC AWAKE2 and the John Adams Institute grants ST/T001933/1, ST/X005518/1 and UKRI1888.} The simulation work was supported through computational resources and services provided by Advanced Research Computing at the University of Michigan, Ann Arbor.
\end{acknowledgments}




\bibliography{apssamp}

\end{document}